\documentclass[pra,preprint,superscriptaddress,floatfix,showkeys,showpacs]{revtex4-1}

\usepackage{amsmath,amsfonts,amssymb}
\usepackage{graphicx,color}
\allowdisplaybreaks

\def\beq{\begin{equation}}
\def\eeq{\end{equation}}
\def\beqn{\begin{eqnarray}}
\def\eeqn{\end{eqnarray}}
\def\beqs{\begin{subequations}}
\def\eeqs{\end{subequations}}

\def\C {G}

\begin{document}

\title{Entanglement and correlations in an exactly-solvable model of a Bose-Einstein condensate in a cavity}
\author{Ofir E. Alon}
\email{ofir@research.haifa.ac.il}
\affiliation{Department of Mathematics, University of Haifa, Haifa 3498838, Israel}
\affiliation{Haifa Research Center for Theoretical Physics and Astrophysics, University of Haifa,
Haifa 3498838, Israel}
\author{Lorenz S. Cederbaum}
\affiliation{Theoretische Chemie, Physikalisch-Chemisches Institut,
Universit\"at Heidelberg, Im Neuenheimer Feld 229, D-69120 Heidelberg, Germany}

\begin{abstract}
An exactly solvable model of
a trapped interacting Bose-Einstein condensate (BEC) coupled in the dipole approximation 
to a quantized light mode in a cavity is presented.
The model can be seen as a generalization of the harmonic-interaction model for a trapped BEC coupled to a bosonic bath.
After obtaining the ground-state energy and wavefunction in closed form,
we focus on computing the correlations in the system.
The reduced one-particle density matrices of the bosons
and the cavity are constructed and diagonalized analytically,
and the von Neumann entanglement entropy of the BEC and the cavity is
also expressed explicitly
as a function of the number and mass of the bosons,
frequencies of the trap and cavity, and the cavity-boson coupling strength.
The results allow one to study the impact of the cavity on the
bosons and vice versa on an equal footing.
As an application we investigate a specific case
of basic interest for itself, namely,
non-interacting bosons in a cavity.
We find that both the bosons and the cavity develop
correlations in a complementary manner while increasing the coupling between them.
Whereas the cavity wavepacket broadens in Fock space,
the BEC density saturates in real space.
On the other hand, while the cavity depletion saturates,
and hence does the BEC-cavity entanglement entropy,
the BEC becomes strongly correlated and eventually increasingly fragmented.
The latter phenomenon implies single-trap fragmentation of otherwise ideal bosons,
where their induced long-range interaction is mediated by the cavity.
Finally, as a complimentary investigation,
the mean-field equations for the BEC-cavity system are solved analytically as well,
and the breakdown of mean-field theory for the cavity and the bosons with increasing coupling
is discussed.
Further applications are envisaged.
\end{abstract}

\maketitle 

\section{Introduction}\label{SEC1}

Atoms and molecules in optical cavities have drawn much attention
in the vast context of how the cavity modifies their bare properties \cite{cav1,cav2,cav3,cav4,cav5,cav6,cav7}.
The exploration of
Bose-Einstein condensates (BECs) in 
driven-dissipative cavities has transcended
from how coupling to the cavity changes their density profiles 
to which correlations and phases are induced within them,
being interacting many-particle systems
\cite{cavb1,cavb2,cavb3,cavb4,cavb5,cavb6,cavb7}.
A fundamental and practical goal is to understand
what governs the entanglement
between the many-particle system and the quantum light,
aiming at controlling and subsequently engineering
desired properties of both,
the many-particle system and the light \cite{cav_ent1,cav_ent2}.

In the present work we aim at investigating
the properties of a BEC in a cavity
from the point of view of an exactly and analytically solvable many-body model.
Furthermore,
we wish to have all quantities,
the many-body properties of the BEC,
correlations in the cavity,
and the entanglement between the BEC and the cavity,
be expressed in closed form
as functions of the number and mass of the bosons,
strength of boson-boson interaction,
frequencies of the trap holding the bosons and cavity mode,
and the strength of the BEC-cavity coupling.
Explicitly, we will show how coupling of the BEC and the cavity
governs the reduced one-particle density matrices and their eigenvalues
for the bosons and cavity alike,
the decay of Galuber's first-order spatial correlation function of the BEC,
and the Schmidt decomposition of the BEC-cavity wavefunction, i.e.,
the von Neumann entanglement entropy.

A good candidate for such a solvable many-body model, as we shall also see below,
draws from the harmonic-interaction model,
in which the bosons are held in an harmonic trap
and interact by harmonic forces.
Harmonic-interaction models have been used,
among others,
to investigate and emulate correlations
and entanglement in bosonic and fermionic systems \cite{Robinson_1977,HIM_COHEN,
HIM1,HIM2,HIM3,HIM4,HIM5,HIM6,HIM7,HIM8,HIM9}.
Another valuable application of harmonic-interaction models
is for benchmarking numerical approaches
to the many-body problem \cite{HIM_BENCH_B,HIM_BENCH_F,HIM_BENCH_BB,HIM_BENCH_CC}.

The structure of the paper is as follows.
In Sec.~\ref{SEC2}, the model is presented and its solution worked out.
In Sec.~\ref{SEC3}, the impact of the cavity on correlation properties of the BEC are derived and investigated,
and in Sec.~\ref{SEC4} the other way around.
Interestingly, although the bosons are represented by a spatial coordinate
and the light is represented in Fock space,
analogies between the structures of their properties and correlations can be drawn.
The von Neumann entanglement entropy is computed and discussed in Sec.~\ref{SEC5}.
In Sec.~\ref{SEC6}, an application of the theory for non-interacting bosons coupled to a cavity is presented.
Finally, concluding remarks including an outlook are collected in Sec.~\ref{SEC7}.
The appendix provides a complementary brief investigation
of the BEC-cavity system at the mean-field level of theory.

\section{The model and its solution}\label{SEC2}

Consider $N$ bosons of mass $m$ held in a trap $V(x)$ and interacting by two-body forces $W(x-x')$ in one spatial dimension.
The bosons are furthermore put in a cavity and coupled to a single-mode quantized light of frequency $\omega_C$
through their dipole moment.
Apart from that,
the bosons are said to be structureless.
These set up the following generic many-body model Hamiltonian
\beqn\label{HAM_BC}
& &
H = H_B + H_C + H_{BC}, \qquad \qquad
H_B = \sum_{j=1}^{N} \left[ -\frac{1}{2m} \frac{\partial^2}{\partial x_j^2} + 
V(x_j) \right] + \sum_{1 \le j < k}^{N} W(x_j-x_k), \nonumber \\
& & 
H_C = \omega_C \left(\hat a^\dag \hat a + \frac{1}{2} \right), \qquad \qquad
H_{BC} = g \left(\hat a^\dag + \hat a\right) \sum_{j=1}^N x_j +
\frac{1}{\omega_C}g^2 \left(\sum_{j=1}^N x_j\right)^2. 
\eeqn
In (\ref{HAM_BC}), $H_B$ is the many-boson Hamiltonian with one-body trapping and two-body interaction terms,
$H_C$ the single-mode harmonic oscillator of the cavity,
and $H_{BC}$ is the coupling Hamiltonian of the bosons and cavity including
the quadratic dipole self-interaction term,
where $g$ is the coupling strength between the bosons and the cavity light.

We begin with
the standard transformation of the cavity mode from second to first quantization
\beqn\label{12Q}
& &
q = \sqrt{\frac{1}{2m\omega_C}} \left(\hat a^\dag + \hat a\right), \qquad
i\frac{d}{\partial q} = i\sqrt{\frac{m\omega_C}{2}} \left(\hat a^\dag - \hat a\right),
\eeqn
where a fictitious mass is introduced in (\ref{12Q}).
This mass can be arbitrarily chosen,
and we thus set it for convenience as the mass $m$ of a boson.

Our aim is to develop and then employ a solvable many-particle model for the BEC-cavity system.
To this end, we consider bosons interacting by harmonic forces 
of strength $\lambda_B$ and trapped in a harmonic potential of frequency $\omega_B$.
This is the harmonic-interaction model for a BEC \cite{HIM_COHEN,HIM1,HIM2,HIM3},
$H_B = \sum_{j=1}^{N} \left( -\frac{1}{2m} \frac{\partial^2}{\partial x_j^2} + 
\frac{1}{2} m\omega_B^2 x_j^2 \right) + \lambda_B \sum_{1 \le j < k}^{N} \left(x_j-x_k\right)^2$.
The harmonic-interaction model is solved using
Jacoby coordinates,
$Q_k = \frac{1}{\sqrt{k(k+1)}} \sum_{j=1}^{k} (x_{k+1}-x_j), k=1,2,\ldots, N-1$
which are relative coordinates
and
$Q_N = \frac{1}{\sqrt{N}} \sum_{j=1}^{N} x_j$
which is the center-of-mass coordinate.
Correspondingly,
diagonalizing the harmonic-interaction model $H_B$
leads to $N-1$ oscillators with frequency $\Omega_B = \sqrt{\omega_B^2 + \frac{2\lambda_BN}{m}}$
and one center-of-mass oscillator with frequency $\omega_B$.

Combining the above for the bosons and the cavity mode,
the intermediate result is
\beqn\label{HAM_BC_JAC1}
& & 
H = \sum_{k=1}^{N-1} \left( -\frac{1}{2m} \frac{\partial^2}{\partial Q_k^2} +
\frac{1}{2} m {\Omega_B}^2 Q_k^2 \right) +
\left[-\frac{1}{2m} \frac{\partial^2}{\partial Q_N^2} + \frac{1}{2} m \left(\omega_B^2+\frac{\C^2}{\omega_C^2}\right) Q_N^2\right] + \nonumber \\
& &
+ \left(-\frac{1}{2m} \frac{\partial^2}{\partial q^2} + \frac{1}{2} m \omega_C^2 q^2\right) +
m \C q Q_N, \qquad \C=g \sqrt{\frac{2\omega_CN}{m}}.
\eeqn
See \cite{FERM_HO} for the system with fermions and Coulomb interaction
in harmonic trap.
The case of a single harmonic degree-of-freedom coupled to the cavity mode
has been treated in \cite{NIM_HO}.

The frequencies' matrix of (\ref{HAM_BC_JAC1}) couples the center-of-mass of the bosons $Q_N$ and the cavity degree-of-freedom $q$
and reads
$\begin{pmatrix}
\omega_B^2 + \frac{\C^2}{\omega_C^2} & \C \cr
\C & \omega_C^2 \cr
\end{pmatrix}$,
where $\C=g \sqrt{\frac{2\omega_CN}{m}}$ is hereafter
referred to as the coupling parameter.
Naturally, the coupling $\C$ governs much of the physics in the BEC-cavity system and is discussed extensively below.
Diagonalization the frequencies' matrix, see \cite{FERM_HO,NIM_HO},
one readily finds the two roots
\beqn\label{BEC_CAV_FREQ}
& & \Omega_{\pm} = \sqrt{\frac{1}{2}\left\{\left(\omega_B^2 + \frac{\C^2}{\omega_C^2} + \omega_C^2\right) \pm 
\sqrt{\left(\omega_B^2 + \frac{\C^2}{\omega_C^2} + \omega_C^2\right)^2 - 4\omega_B^2\omega_C^2}\right\}}. \
\eeqn
The frequencies $\Omega_{\pm}$
do not depend on the sign of the coupling parameter $\C$.
We hence consider without loss of generality positive coupling parameters from now on. 
The two frequencies (\ref{BEC_CAV_FREQ}) are always positive,
implying the boundedness of the solution due to the coupled $Q_N$ and $q$ degrees-of-freedom, see below.
$\Omega_+$ increases with enlarging the
coupling parameter $\C$ while
$\Omega_-$ decreases.
The resulting coupled center-of-mass Jacoby coordinates read
\beqn\label{Qpm}
& & Q_+ =
\frac{\C Q_N + D q}{\sqrt{\C^2+D^2}}, \quad
Q_- =
\frac{-D Q_N + \C q}{\sqrt{\C^2+D^2}}, \nonumber \\
& &
D=\frac{1}{2}\left\{\sqrt{\left(\omega_B^2 + \frac{\C^2}{\omega_C^2} + \omega_C^2\right)^2-4\omega_B^2\omega_C^2} -
\left(\omega_B^2 + \frac{\C^2}{\omega_C^2} -\omega_C^2\right)\right\}. \
\eeqn
It is instructive to examine limiting cases.
Two obvious limits are for very weak or very strong coupling.
In the first,
when $\frac{\C}{\omega_B\omega_C} \ll 1$,
one has
$\Omega_+ \to \omega_C$, $Q_+ \to Q_N$ and 
$\Omega_- \to \omega_B$, $Q_- \to q$ for $\omega_B > \omega_C$,
and the other way around for $\omega_B < \omega_C$, see (\ref{Qpm}).
In the second limit,
when, $\frac{\C}{\omega_B\omega_C} \gg 1$,
one finds to leading order in $\C$ and irrespective of
$\omega_B$ and $\omega_C$
the pace of increase $\Omega_+ \to \frac{\C}{\omega_C}$
and decrease $\Omega_- \to \frac{\omega_B\omega_C^2}{\C}$ of the roots,
and for later use $D \to \omega_C^2$.
Interestingly, since $Q_+ \to Q_N$ and $Q_- \to q$ hold for $\frac{\C}{\omega_B\omega_C} \gg 1$,
one might conclude that the BEC and the cavity become
decoupled.
As we shall see below, this is not the case.
Clearly, the coupled center-of-mass Jacoby coordinates $Q_\pm$ and more so their frequencies $\Omega_{\pm}$
are essential in driving
the correlation properties of the BEC-cavity system.

After completing the separation of variables in (\ref{HAM_BC_JAC1}),
we get $N+1$ decoupled harmonic oscillators.
The ground-state energy is hence
\beqn\label{GS_BEC_CAV}
& &
E = \frac{1}{2}\left[\left(N-1\right)\Omega_B + \Omega_+ + \Omega_-\right].
\eeqn
The energy (\ref{GS_BEC_CAV}) is bound from below by $\frac{1}{2}\left(\omega_B+\omega_C\right)$
and unbound from above.
The former occurs in the absence of coupling
between the bosons and the cavity,
and when the repulsion between the bosons
pushes the frequency of the relative degrees-of-freedom $\Omega_B$
to $0^{+}$, i.e., to the verge of being bound.
The unboundedness of (\ref{GS_BEC_CAV}) from above
reflects the nature of harmonic interaction between 
bosons in the harmonic-interaction model
and, independently, the ever increasing upper root $\Omega_+$
as the coupling $\C$ is made stronger and stronger.

The wavefunction of the BEC-cavity system
corresponding to the energy (\ref{GS_BEC_CAV})
reads in terms of the Jacoby coordinates
\beqn\label{BC_WF_JAC}
& &
\Psi(Q_1,\ldots,Q_{N-1},Q_+,Q_-) = 
\left(\frac{m\Omega_B}{\pi}\right)^{\frac{N-1}{4}}
\left(\frac{m\Omega_+}{\pi}\right)^{\frac{1}{4}}
\left(\frac{m\Omega_-}{\pi}\right)^{\frac{1}{4}}
\times \nonumber \\
& & \times
e^{-\frac{1}{2}m\left(\Omega_B \sum_{k=1}^{N-1} Q_k^2 +
\Omega_+ Q_+^2 + \Omega_- Q_-^2\right)}. \
\eeqn
It is useful to transform the BEC-cavity wavefunction to the laboratory frame and
express (\ref{BC_WF_JAC}) in Cartesian coordinates
\beqn\label{BC_WF_CAR}
& & \Psi(x_{1},\ldots,x_{N},q) = 
\left(\frac{m\Omega_B}{\pi}\right)^{\frac{N-1}{4}}
\left(\frac{m\Omega_+}{\pi}\right)^{\frac{1}{4}}
\left(\frac{m\Omega_-}{\pi}\right)^{\frac{1}{4}}
\times \nonumber \\
& & \times
e^{-\frac{\alpha_B}{2} \sum_{j=1}^{N} x_{j}^2 - \beta_B \sum_{1\le j<k}^{N} x_{j}x_{k}} \,
e^{-\frac{\alpha_C}{2} q^2} \,
e^{+\gamma  \sum_{j=1}^{N} x_{j} q}, \
\eeqn
where the four coefficients are
\beqn\label{BC_WF_CAR_PARAMs}
& &
\alpha_B = m\Omega_B+\beta_B, \qquad
\beta_B = \frac{m}{N}\left\{-\Omega_B + \frac{\C^2\Omega_+ + D^2\Omega_-}{\C^2+D^2}\right\}, \nonumber \\
& &
\alpha_C = m \frac{D^2\Omega_+ + \C^2\Omega_-}{\C^2+D^2}, \qquad
\gamma = \frac{m}{\sqrt{N}} \frac{\C D}{\C^2+D^2} \left(\Omega_- - \Omega_+\right). \
\eeqn
All coefficients of the wavefunction (\ref{BC_WF_CAR}) are seen to depend on the BEC-cavity coupling $\C$ --
these are the one-body and two-body parts of the bosons, $\alpha_B$ and $\beta_B$, respectively,
the one-body part of the cavity, $\alpha_C$,
and, naturally,
the two-body BEC-cavity part $\gamma$.
The structure of (\ref{BC_WF_CAR}) resembles that of the wavefunction of a mixture of
bosons \cite{HIM_IJQC,HIM_MIX_ENTANGLE,BB_HIM},
with $N$ bosons of one kind and a single particle of another kind,
albeit the coefficients are more complicated.
The model 
(\ref{HAM_BC_JAC1})
can be seen as
a generalization of the harmonic-interaction model for a trapped BEC to include coupling to a bosonic bath.
Techniques developed for the harmonic-interaction model for
trapped bosonic mixtures \cite{BB_HIM,BB_2022,BB_2023} can be utilized here.
We now proceed to
investigate the impact of the coupling between the BEC and the cavity on their respective
correlation properties.

\section{Impact of the cavity on the BEC}\label{SEC3}

To quantify the impact of the cavity in combination with boson-boson interaction on the BEC
one needs to precisely fold the cavity and bosonic degrees-of-freedom.
Starting from the wavefunction (\ref{BC_WF_CAR}) and
integrating out $N-1$ bosons and the cavity mode following \cite{BB_HIM},
one finds the reduced one-particle density of the BEC
\beqn\label{1RDM_BEC}
& &
\rho_B(x,x') = N \left(\frac{\alpha_B+C_{1,0}}{\pi}\right)^{\frac{1}{2}}
e^{- \frac{\alpha_B+C_{1,0}}{4}\left(x+x'\right)^2}
e^{-\frac{\alpha_B}{4}\left(x-x'\right)^2}, \nonumber \\
& &
\alpha_B+C_{1,0} = (\alpha_B-\beta_B)
\frac{[(\alpha_B-\beta_B)+N\beta_B] \alpha_C - \gamma^2N}
{[(\alpha_B-\beta_B)+(N-1)\beta_B] \alpha_C - \gamma^2(N-1)}, \
\eeqn
where the coefficient $C_{1,0}$ governs the correlations in the BEC.
$C_{1,0}$ depends on both
the boson-boson interaction and on
the coupling to the cavity.
In the absence of coupling, $\C=0$, and in the absence
of boson-boson interaction, $\lambda_B=0$,
the coefficient $C_{1,0}$ vanishes and $\rho_B(x,x')$
becomes the reduced one-particle density matrix of fully condensed non-interacting bosons.
Otherwise, $C_{1,0}$ is negative and correlations set in, see below.
The density of the BEC is given by the diagonal of the reduced one-particle density,
$\rho_B(x) \equiv \rho_B(x,x) = N \left(\frac{\alpha_B+C_{1,0}}{\pi}\right)^{\frac{1}{2}}
e^{-\left(\alpha_B+C_{1,0}\right) x^2}$.
The width of the density is
$\frac{1}{\sqrt{\alpha_B+C_{1,0}}}$
and governed by the boson-boson interaction and BEC-cavity coupling.
The impact of the cavity on the density of the BEC
is exemplified in Sec.~\ref{SEC6}.

The eigenstates and eigenvalues of the reduced one-particle density matrix provide
important quantities and are known as natural orbitals and their occupation numbers.
They quantify how the coupling to the cavity in combination with the boson-boson
interaction govern the depletion and subsequently the fragmentation of the BEC.
Fragmentation of BECs,
i.e., the macroscopic occupation of more than a single eigenvalue of the reduced 
one-particle density matrix has been investigated intensively \cite{FRAG1,FRAG2,FRAG3,FRAG4,FRAG5,FRAG6}. 
It is possible to diagonalize the reduced one-particle density matrix using Mehler's formula, see, e.g., \cite{Robinson_1977,HIM5,BB_2023}.
The final result reads
\beqn\label{1RDM_BEC_DIAG}
& & \rho_B(x,x') = N \sum_{n=0}^{\infty} \left(1-\rho_B\right) \rho_B^n \xi_n(x) \xi_N^\ast(x'), \nonumber \\
& & \rho_B = \frac{{\mathcal W}_B-1}{{\mathcal W}_B+1},
\qquad {\mathcal W}_B = \sqrt{\frac{\alpha_B}{\alpha_B+C_{1,0}}}, \
\eeqn
where $N\left(1-\rho_B\right)\rho_B^n, n=0,1,2,3,\ldots$
are the natural occupation numbers.
The sum of all occupation numbers is the number of bosons $N$.
The lowest occupation number per particle
provides the so called condensate fraction $\left(1-\rho_B\right)$
and $\rho_B$ is hence the depleted fraction.
The impact of the cavity on the depleted fraction of the BEC
is investigated in Sec.~\ref{SEC6}.
The natural orbitals are
$\xi_n(x) = \frac{1}{\sqrt{2^n n!}}\left(\frac{s_B}{\pi}\right)^{\frac{1}{4}}H_n(\sqrt{s_B}x) e^{-\frac{1}{2}s_Bx^2}$,
with $H_n(x)$ the Hermite polynomials and $s_B=\sqrt{\alpha_B\left(\alpha_B+C_{1,0}\right)}$
the scaling due to boson-boson interaction $\lambda_B$ and couping $\C$ to the cavity. 

Another way to describe the correlations in the BEC is spatially.
For this, combining the reduced one-particle density matrix and
the density of the bosons is required.
Using (\ref{1RDM_BEC_DIAG}) and its diagonal,
Glauber's normalized first-order correlation function \cite{GLA1,GLA2} is readily obtained explicitly and takes on a simple form,
\beqn\label{g1_BEC}
g_B(x,x') = \frac{\rho_B(x,x')}{\sqrt{\rho_B(x)\rho_B(x')}} = e^{-\frac{-C_{1,0}}{4}\left(x-x'\right)^2}.
\eeqn
In the absence of interaction $\lambda_B$ and coupling $\C$,
$g_B(x,x')=1$ as is expected for non-interacting bosons in the absence of a cavity.
In this case the bosons exhibit full spatial first-order coherence
and the reduced one-particle density matrix is in a product form.
When interaction between the bosons or coupling to the cavity set in,
$\rho_B(x,x')$ departs from a product form and consequently
the first-order correlation function (\ref{g1_BEC}) decays along the off diagonal.
This means that the further away bosons in two positions $x$ and $x'$ are from one another,
the less they are spatially coherent.

The length scale for the decay of $g_B(x,x')$
is not the only length scale describing the BEC and alone is
insufficient to assess the build up of first-order correlations.
To quantify further the correlations accompanying the off-diagonal decay of (\ref{g1_BEC}),
it is instrumental to compare the length scale of this decay,
given by $\sqrt{\frac{4}{-C_{1,0}}}$,
to the size of the BEC
or, more precisely, 
the length scale of the decay of the diagonal term in
$\rho_B(x,x')$ (\ref{1RDM_BEC}),
given by $\sqrt{\frac{4}{\alpha_B+C_{1,0}}}$.
Thus, we introduce the dimensionless coherence length as the ratio of the two decay length scales,
\beqn\label{BEC_LB}
l_B = \sqrt{\frac{\alpha_B+C_{1,0}}{-C_{1,0}}} = \frac{1}{\sqrt{\left(\frac{1+\rho_B}{1-\rho_B}\right)^2-1}}. \
\eeqn
We stress that the coherence length $l_B$
is a dimensionless quantity.

It is senseful to define the boarder between weak and strong correlations at $l_B=1$.
Indeed, for vanishing correlations 
the bosons remain spatially coherent over long distances
and $l_B \to \infty$,
and for very strong correlations 
the bosons lose spatial coherence over short distances and $l_B \to 0^+$.
Furthermore, the coherence length $l_B$ is neatly related to the depletion $\rho_B$,
as the second expression in (\ref{BEC_LB}) shows.
As expected,
diminishing the depletion the coherence length increases unboundedly and increasing the depletion
the coherence length shrinks towards zero.
Finally,
solving (\ref{BEC_LB}) for $l_B=1$
one gets $\rho_B=\frac{\sqrt{2}+1}{\sqrt{2}-1} \approx 17.15\%$
for the depleted fraction at the weak-to-strong correlations border.

Sec.~\ref{SEC6} presents an investigation of the depletion, fragmentation, Glauber's first-order spatial correlation function,
and dimensionless coherence length of the BEC in the cavity.

\section{Impact of the BEC on the cavity}\label{SEC4}

Just as the cavity impacts the BEC,
the coupling of the BEC to the cavity induces correlations in the latter.
To quantify the correlations,
we turn now to computing properties of the cavity mode.
This requires precisely folding all bosonic degrees-of-freedom.
Starting from the wavefunction (\ref{BC_WF_CAR}) and
integrating out the $N$ bosons following \cite{BB_HIM},
one obtains the reduced one-particle density matrix of the cavity mode
\beqn\label{1RDM_CAV}
& &
\rho_C(q,q') = \left(\frac{\alpha_C+C'_{0,1}}{\pi}\right)^{\frac{1}{2}}
e^{- \frac{\alpha_C+C'_{0,1}}{4}\left(q+q'\right)^2}
e^{-\frac{\alpha_C}{4}\left(q-q'\right)^2}, \qquad
C'_{0,1} = -\gamma^2\frac{N}{(\alpha_B-\beta_B)+N\beta_B}. \nonumber \\ \
\eeqn
Perhaps, the first matter to notice is that the cavity $\rho_C(q,q')$
and BEC $\rho_B(x,x')$ reduced one-particle density matrices possess equivalent structures.
This is not surprising, and draws from the resemblance of the wavefunction (\ref{BC_WF_CAR})
to that of a mixture of two kinds of identical bosons. 
The coefficient $C'_{0,1}$ governs the correlations induced
in the cavity mode and depends on the coupling parameter $\C$
but not on the boson-boson interaction,
reflecting the fact that the
relative-motion degrees-of-freedom of
the BEC, $Q_k, k=1,\ldots,N-1$, are not coupled to the cavity.
Furthermore, comparing the cavity coefficient $C'_{0,1}$ to
the BEC coefficient $C_{1,0}$ in the respective reduced density matrices,
the former is found to be slightly simpler.
This is because there is a single cavity mode on the one hand,
and $N$ bosons on the other hand.
It should be stressed that the representation (\ref{1RDM_CAV}) 
emerges when translating the cavity mode from second to first quantization,
and is not to be interpreted for spatial properties of the cavity mode.
We are
interested in the structure of $\rho_C(q,q')$ and its eigenvalues and their consequences.

In the same manner that the reduced one-particle density matrix of $N$ bosons
defines the degree of fragmentation of a BEC \cite{FRAG1,FRAG2,FRAG3,FRAG4,FRAG5,FRAG6},
even down to $N=2$ interacting bosons \cite{2B_FRAG},
we may ask and answer 
the analogous question regarding the single cavity mode,
owing to its coupling to the BEC.
Indeed,
it is possible to exactly diagonalize the reduced one-particle density matrix of the cavity mode using
Mehler's formula \cite{Robinson_1977,HIM5,BB_2023}.
Formally,
this will provide us with the natural occupation numbers and orbitals,
and quantify how the coupling of
the single cavity mode to the bosons alters the correlation properties of the cavity.
Hence, the final result is analogous to (\ref{1RDM_BEC_DIAG}) and reads
\beqn\label{1RDM_CAV_DIAG}
& & \rho_C(q,q') = \sum_{n=0}^{\infty} \left(1-\rho_C\right) \rho_C^n \phi_n(q) \phi_n^\ast(q'), \nonumber \\
& & \rho_C = \frac{{\mathcal W}_C-1}{{\mathcal W}_C-1}, \qquad
{\mathcal W}_C = \sqrt{\frac{\alpha_C}{\alpha_C+C'_{0,1}}}, \
\eeqn
where $\left(1-\rho_C\right) \rho_C^n, n=0,1,2,3,\ldots$
are the occupation numbers.
The latter suggest that $\rho_C$ can be interpreted as the depleted fraction of the cavity mode.
Of course, the occupation numbers sum up to one.
For completeness,
the natural orbitals of the cavity are
$\phi_n(q) = \frac{1}{\sqrt{2^n n!}}\left(\frac{s_C}{\pi}\right)^{\frac{1}{4}}H_n(\sqrt{s_C}q) e^{-\frac{1}{2}s_Cq^2}$,
where
$s_C=\sqrt{\alpha_C\left(\alpha_C+C'_{0,1}\right)}$ is
the scaling due to coupling to the BEC. 

As noted above,
the structures of the BEC and cavity reduced one-particle density matrices
(\ref{1RDM_BEC}) and (\ref{1RDM_CAV}) are equivalent.
The real-space coordinate $x,x'$ in the former and the first-quantization coordinate $q,q'$
in the latter appear in the same functional form.
Thus, we may draw further analogies.
The diagonal of the reduced one-particle density matrix can be defined,
$\rho_C(q) \equiv \rho_C(q,q) = \left(\frac{\alpha_C+C'_{0,1}}{\pi}\right)^{\frac{1}{2}}
e^{-\left(\alpha_C+C'_{0,1}\right) q^2}$,
and called a density.
The quantity $\rho_C(q)$
is not a real-space density but has similar properties.
For instance,
$\frac{1}{\sqrt{\alpha_C+C'_{0,1}}}$
is its width.
Then, a first-order correlation function for the cavity mode may be prescribed,
$g_C(q,q') = \frac{\rho_C(q,q')}{\sqrt{\rho_C(q)\rho_C(q')}} = e^{-\frac{-C'_{0,1}}{4}\left(q-q'\right)^2}$.
In the absence of coupling between the BEC and the cavity
$g_C(q,q')=1$ is flat,
as one would obtain for a non-interacting system.
Otherwise, $g_C(q,q')$ decays along its off diagonal.
Next, analogously to $l_B$ of the bosons (\ref{BEC_LB}),
we may define the cavity
dimensionless coherence length $l_C$ and express it using the depletion $\rho_C$.
The final result reads
\beqn\label{VAC_LC}
l_C = \sqrt{\frac{\alpha_C+C'_{0,1}}{-C'_{0,1}}} = \frac{1}{\sqrt{\left(\frac{1+\rho_C}{1-\rho_C}\right)^2-1}}. \
\eeqn
We stress again that $l_C$ is not a physical length characterizing spatial coherence or spatial correlations induced in
the cavity by the BEC.
Rather, it is a dimensional parameter that compares the off-diagonal decay of $g_C(q,q')$
with the diagonal decay of $\rho_C(q,q')$.
$l_C$ diverges in the no coupling limit and goes to zero in the strong coupling limit,
and thus mimics the physical behavior of a real-space coherence length,
like $l_B$ of the bosons.

Finally, one can collect the whole set of occupation numbers (\ref{1RDM_CAV_DIAG}) and
define the entanglement entropy of the cavity reduced one-particle density matrix,
\beqn\label{1RDM_C_ENT}
& &
{\mathcal S}_C = - \sum_{n=0}^\infty \left(1-\rho_C\right) \rho_C^n \ln[\left(1-\rho_C\right) \rho_C^n] =
-\left[\ln(1-\rho_C) + \frac{\rho_C\ln(\rho_C)}{1-\rho_C}\right].
\eeqn
The entanglement entropy ${\mathcal S}_C$
is a monotonously increasing function of the depletion $\rho_C$ and is not bound from above.
It varies in an opposite manner to the coherence length $l_C$
which is a monotonously decreasing function of the depletion $\rho_C$ and is not bound from above.
In turn,
quantifying the correlations in the cavity
either by the depletion $\rho_C$,
the dimensionless coherence length $l_C$,
or the entanglement entropy ${\mathcal S}_C$
are parallel.
Finally, 
that expression (\ref{1RDM_C_ENT})
is also the entanglement entropy of the BEC-cavity system
is discussed and expanded in the next Sec.~\ref{SEC5}.

The application presented in Sec.~\ref{SEC6}
explores the depletion, fragmentation,
and dimensionless coherence length of the cavity mode,
when coupled to the BEC.

\section{Entanglement between the BEC and the cavity mode}\label{SEC5}

The final piece of information resulting from the coupling 
of the BEC and the cavity
is the von Neumann entanglement entropy between them. 
Returning to a representation of the wavefunction in terms of the BEC Jacoby coordinates $Q_k, k=1,\dots,N$
and the cavity-mode coordinate $q$,
one can perform the Schmidt decomposition of the BEC-cavity wavefunction.
Recall that only the center-of-mass coordinate $Q_N$ is coupled to the cavity coordinate $q$.
The weights of this decomposition characterize to what extent
the two subsystems are entangled to each other.

Substituting the coupled center-of-mass Jacoby coordinates (\ref{Qpm}) into 
the diagonalized wavefunction (\ref{BC_WF_JAC}) one finds
\beqn\label{BC_WF_ENT}
& &
\Psi(Q_1,\ldots,Q_{N-1},Q_N,q) = 
\left(\frac{m\Omega_B}{\pi}\right)^{\frac{N-1}{4}}
\left(\frac{m\Omega_+}{\pi}\right)^{\frac{1}{4}}
\left(\frac{m\Omega_-}{\pi}\right)^{\frac{1}{4}}
e^{-\frac{1}{2}m\Omega_B \sum_{k=1}^{N-1} Q_k^2} \times \nonumber \\
& &
\times
e^{-\frac{1}{2}m\left(\frac{\C^2\Omega_+ + D^2 \Omega_-}{\C^2+D^2}Q_N^2 +
\frac{D^2 \Omega_+ + \C^2\Omega_-}{\C^2+D^2} q^2\right)}
e^{-m\frac{\C D}{\C^2+D^2}\left(\Omega_+ - \Omega_-\right) Q_N q}, \
\eeqn
where, to remind, $\C$ is the coupling parameter and $D$ is given in (\ref{Qpm}).
The form (\ref{BC_WF_ENT}) of the wavefunction
is not yet amenable to decomposition via Mehler's formula,
because the two coupled degrees-of-freedom,
$Q_N$ and $q$, are not ``equivalent".
To this end,
a squeeze transformation is required \cite{BB_2022},
i.e., definition of new variables satisfying
$\tilde Q_N^2 \equiv \sigma Q_N^2$ and $\tilde q^2 \equiv \frac{1}{\sigma} q^2$,
where $\sigma=\sqrt{\frac{\C^2\Omega_+ + D^2 \Omega_-}{D^2 \Omega_+ + \C^2\Omega_-}}$.
Another difference between the utilization of Mehler's formula
to diagonalize
the reduced one-particle density matrices (\ref{1RDM_BEC},\ref{1RDM_CAV}) in the previous sections 
and the Schmidt decomposition
to be performed onto the wavefunction (\ref{BC_WF_ENT})
is the normalization of the eigenvalues (weights). 
In the former case the sum of the natural occupation numbers per particle adds up to
one whereas in the latter it is the sum of squares of the Schmidt-decomposition weights which should be one.
Putting all together,
the final result for the Schmidt decomposition is hence
\beqn\label{BC_WF_ENT_SD}
& &
\Psi(Q_1,\ldots,Q_N,q) = \sum_{n=0}^\infty \sqrt{1-\rho_{SD}^2} \rho_{SD}^n \,
\Xi_n(Q_1,\ldots,Q_N) \varphi_n(q), \nonumber \\
& &
\rho_{SD} = \frac{{\mathcal W_{SD}}-1}{{\mathcal W_{SD}}+1}, \qquad
{\mathcal W_{SD}} = \left[\frac{\sqrt{\left(\frac{\Omega_+}{\Omega_-}+1\right)^2+\left(\frac{\C}{D}-\frac{D}{\C}\right)^2\frac{\Omega_+}{\Omega_-}}+\left(\frac{\Omega_+}{\Omega_-}-1\right)}{\sqrt{\left(\frac{\Omega_+}{\Omega_-}+1\right)^2+\left(\frac{\C}{D}-\frac{D}{\C}\right)^2\frac{\Omega_+}{\Omega_-}}-\left(\frac{\Omega_+}{\Omega_-}-1\right)}\right]^{\frac{1}{2}}.
\eeqn
Here,
the Schmidt-decomposition eigenfunctions of the BEC and the cavity are\break\hfill
$\Xi_n(Q_1,\ldots,Q_N) = \left(\frac{m\Omega_B}{\pi}\right)^{\frac{N-1}{4}}
e^{-\frac{1}{2} m \Omega_B \sum_{k=1}^{N-1} Q_k^2}
\frac{\left(-1\right)^n}{\sqrt{2^n n!}} \left(\frac{s\sigma}{\pi}\right)^{\frac{1}{4}}
H_n\left(\sqrt{s\sigma}Q_N\right) e^{-\frac{1}{2}s\sigma Q_N^2}$
and $\varphi_n(q) = \left(\frac{s}{\pi\sigma}\right)^{\frac{1}{4}}
H_n\left(\sqrt{\frac{s}{\sigma}}q\right) e^{-\frac{1}{2}\frac{s}{\sigma} q^2}$,
respectively.
The scaling is $s=m\sqrt{\Omega_+\Omega_-}$ and the squeeze factor $\sigma$ is given above.
Naturally,
the sum of the squares of all Schmidt-decomposition weights in (\ref{BC_WF_ENT_SD}), $\left(1-\rho_{SD}^2\right)\rho_{SD}^{2n}, n=0,1,2,3,\ldots$, adds up to one.
Then, collecting the whole set of squared weighs,
the von Neumann entanglement entropy,
\beqn\label{SD_ENT_ENT}
& &
{\mathcal S} = - \sum_{n=0}^\infty \left(1-\rho_{SD}^2\right) \rho_{SD}^{2n}
\ln[\left(1-\rho_{SD}^2\right) \rho_{SD}^{2n}] =
-\left[\ln(1-\rho_{SD}^2) + \frac{\rho_{SD}^2\ln(\rho_{SD}^2)}{1-\rho_{SD}^2}\right],
\eeqn
between the BEC and the cavity is computed directly.

Recall that there are two equivalent ways to compute the entanglement entropy in a pure bipartite system,
either by performing the Schmidt decomposition of the wavefunction,
or by computing the eigenvalues of the reduced density matrix
of either of its two subsystems \cite{CENT}.
In our case,
$\rho_C(q,q')$ is the reduced density matrix of the cavity mode
and its eigenvalues are computed in (\ref{1RDM_CAV_DIAG}).
Consequently, the reduced-density (\ref{1RDM_C_ENT}) and Schmidt-decomposition (\ref{SD_ENT_ENT}) results are the same,
namely,
$\rho_{SD}^2 = \rho_C$ and
${\mathcal S} = {\mathcal S}_C$.
The von Neumann entanglement entropy
increases monotonously with $\rho_{SD}^2$ (i.e., with the depletion of the cavity),
and we use the latter in accounting for the former.
Sec.~\ref{SEC6} discusses
the entanglement entropy in an application,
where non-interacting bosons
and a cavity mode are coupled.

\section{Application: Non-interacting bosons in a cavity}\label{SEC6}

The many-body model presented in the previous 
sections was solved for interacting bosons.
Below we show that even the simplest case
of which is already very appealing.
Explicitly, we utilize the model to investigate a finite BEC of non-interacting bosons
in harmonic potential
coupled to a cavity mode.

Following the tools derived in the previous sections,
we address and focus on the following questions:
1) How coupling of non-interacting bosons to the cavity 
can make them ``interacting",
and hence to exhibit properties of interacting bosons like depletion, correlations, and even fragmentation?
2) How coupling of the cavity to the bosons can make it correlated?
3) Who is more correlated as a result of their coupling,
the BEC or the cavity mode?
4) What is the entanglement entropy of the BEC-cavity system and how it varies with boson-cavity coupling?
To this end,
we consider $N=100$ non-interacting bosons, $\lambda_B=0$.
When deemed instructive for our discussion,
results for $N=1000$ and $N=10^6$ bosons are presented too, see the Appendix.
The frequency of the relative-motion degrees-of-freedom becomes
the bare oscillator frequency,
$\Omega_{B}=\omega_B$,
and the total energy (\ref{GS_BEC_CAV}) reads then 
$E = \frac{1}{2}\left[\left(N-1\right)\omega_B + \Omega_+ + \Omega_-\right]$.
For simplicity, the mass of a boson is set to $m=1.0$
and its trapping frequency to $\omega_B=1.0$.
The cavity frequency is taken to be $\omega_C=0.1$.
It is convenient to express quantities as a function of the coupling parameter $\C=g \sqrt{\frac{2\omega_CN}{m}}$
in dimensionless units, $\frac{\C}{\omega_B\omega_C}$,
which would allow us to compare different systems to each other.

\begin{figure}[!]
\begin{center}
\hglue -2.5 truecm
\includegraphics[width=0.577\columnwidth,angle=-90]{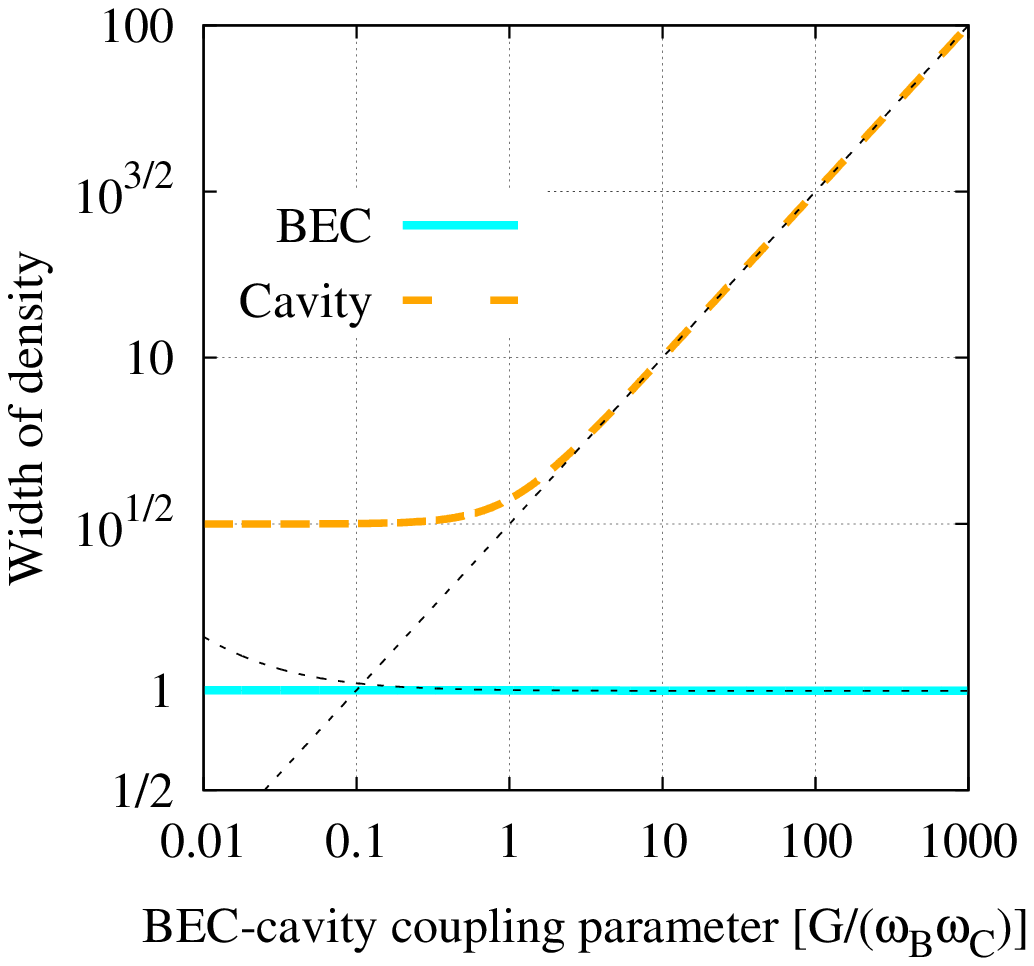}
\end{center}
\vglue 0.15 truecm
\caption{Widths
of the bosons density per particle and the analogously-defined
cavity density.
$N=100$ non-interacting bosons of mass $m=1.0$ in a trap ($\omega_B=1.0$)
are coupled to a single-mode quantum light in a cavity ($\omega_C=0.1$).
The thin dashed lines are the asymptotic behavior of the widths
for large coupling parameters, $\frac{\C}{\omega_B\omega_C} \gg 1$.
See the text and the appendix for further details and discussion.
The quantities shown are dimensionless.}
\label{F1}
\end{figure}

\begin{figure}[!]
\begin{center}
\hglue -1.6 truecm
\includegraphics[width=0.43\columnwidth,angle=-90]{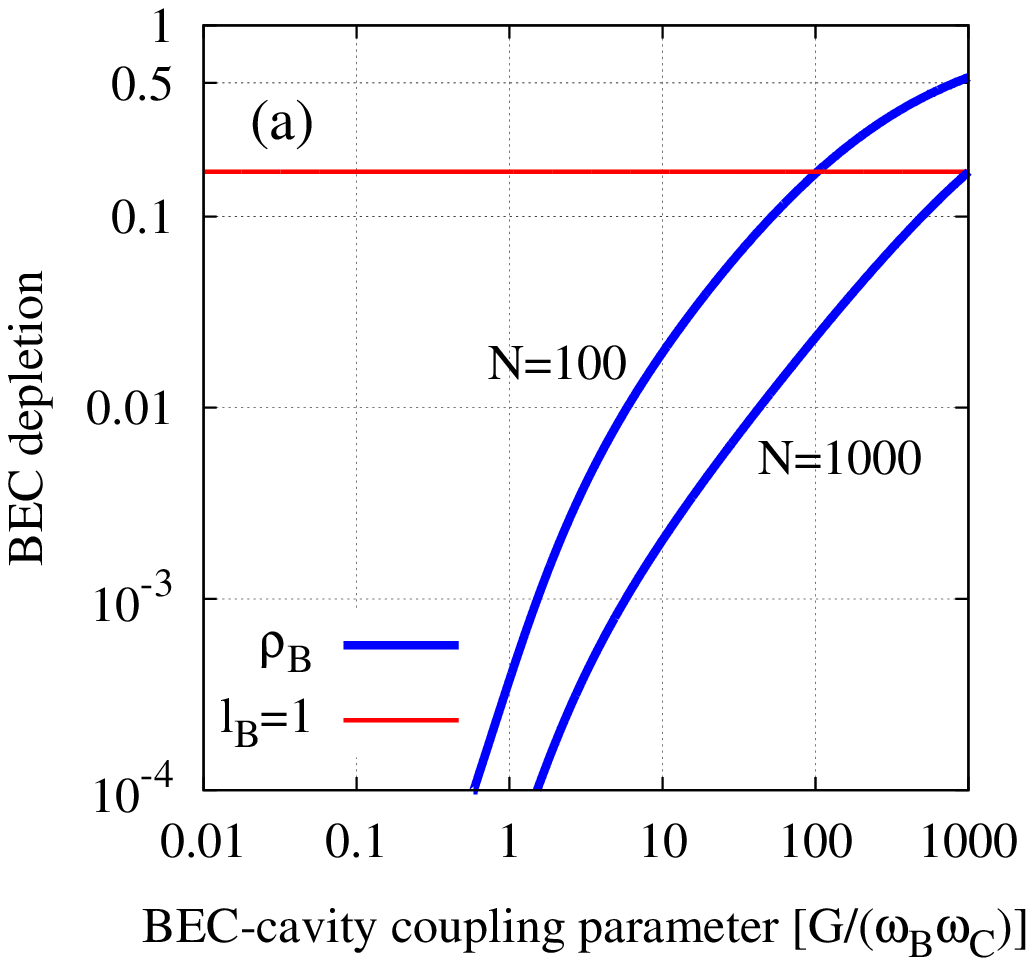}
\hglue -2.8 truecm
\includegraphics[width=0.43\columnwidth,angle=-90]{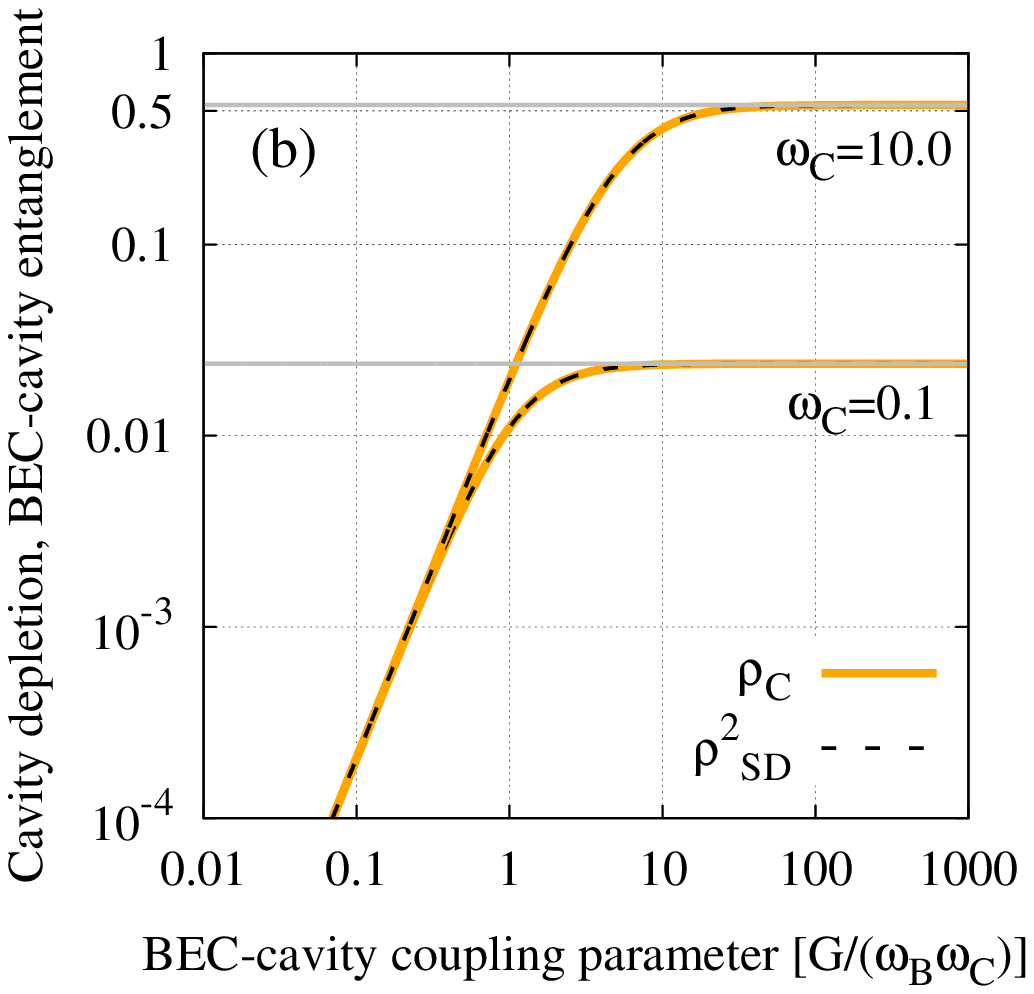}
\end{center}
\vglue 0.15 truecm
\caption{Depletion and entanglement of the BEC and the cavity.
$N=100$ non-interacting bosons of mass $m=1.0$ in a trap ($\omega_B=1.0$)
are coupled to a single-mode quantum light in a cavity ($\omega_C=0.1$).
(a) The depletion $\rho_B$ of the BEC 
and (b) the analogously-computed $\rho_C$ for the cavity mode
are plotted as a function of the coupling parameter $\C$.
The Schmidt-decomposition parameter quantifying the von Neumann 
entanglement entropy of the BEC and the cavity obeys
$\rho_{SD}^2=\rho_C$ and plotted as well.
The horizontal line in (a) marks the value $\rho_B=\frac{\sqrt{2}-1}{\sqrt{2}+1}\approx 0.1715$
for which the dimensionless coherence length of the bosons is $l_B=1$.
The horizontal lines in (b) indicate the asymptotic values of the
cavity depletion (and of the BEC-cavity entanglement entropy)
$\rho_C \to \frac{\sqrt{1+\frac{\omega_C}{\omega_B}}-1}{\sqrt{1+\frac{\omega_C}{\omega_B}}+1}$.
Additional analysis is facilitated in
(a) for $N=1000$ bosons and the same other parameters
and in (b) for $\omega_C=10.0$ and the same other parameters.
See Figs.~\ref{F1} and \ref{F3} and the text for further discussion.
The quantities shown are dimensionless.}
\label{F2}
\end{figure}

\begin{figure}[!]
\begin{center}
\hglue -1.6 truecm
\includegraphics[width=0.32\columnwidth,angle=-90]{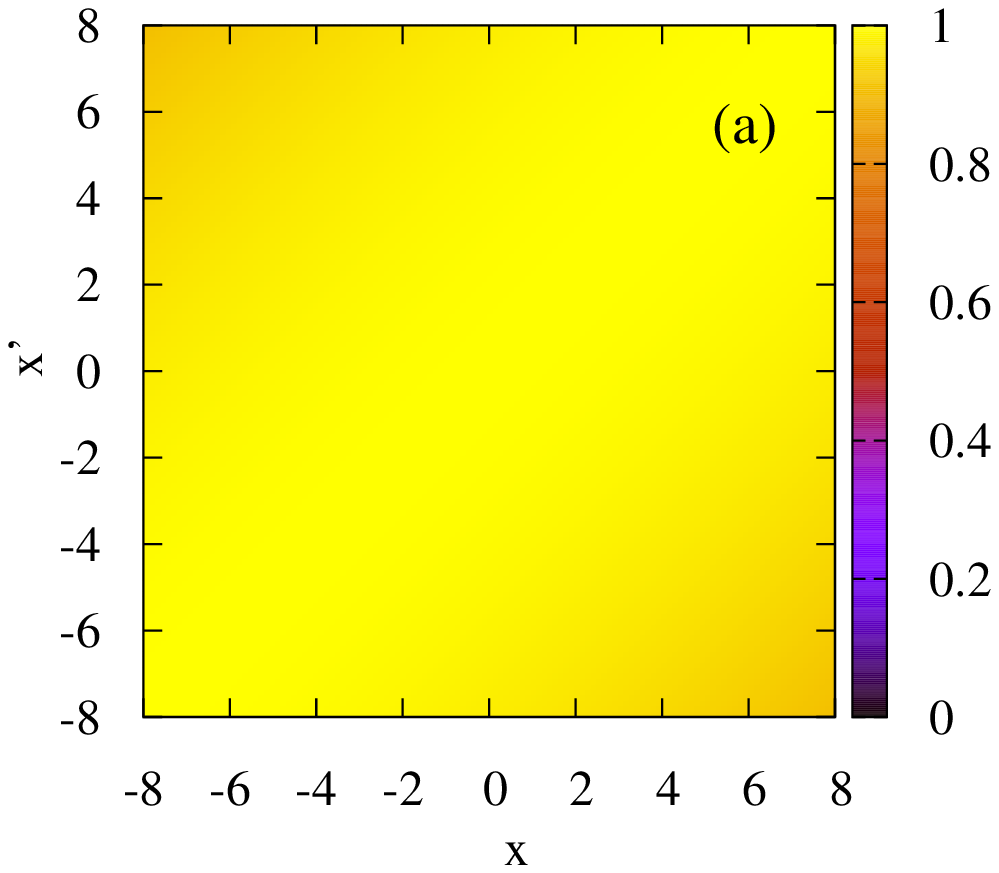}
\hglue -2.4 truecm
\includegraphics[width=0.32\columnwidth,angle=-90]{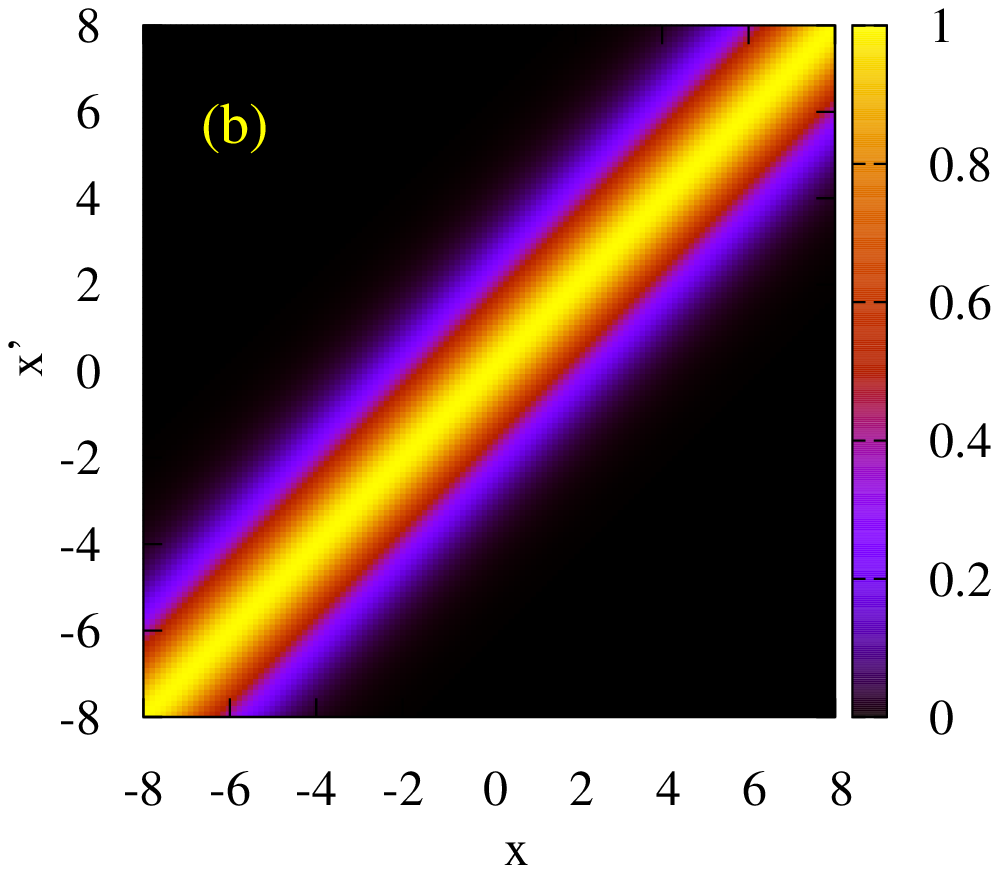}
\hglue -2.4 truecm
\includegraphics[width=0.32\columnwidth,angle=-90]{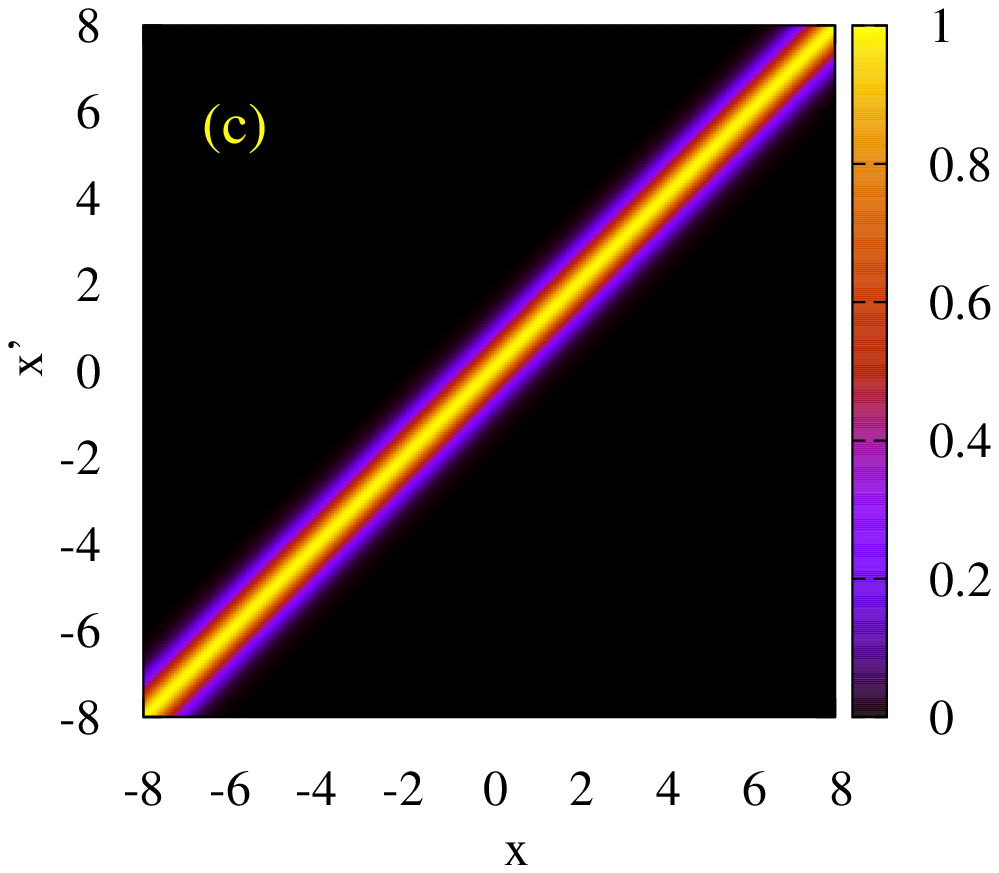}
\end{center}
\vglue 0.15 truecm
\caption{Spatial correlations of the BEC coupled to a cavity.
Shown is Glauber's first-order spatial correlation function of the BEC, $g_B(x,x')$,
for (a) $\frac{\C}{\omega_B\omega_C}=1$,
when the mean-field theory breaks down for the cavity,
and the BEC is weakly depleted ($\rho_B=3.784 \cdot 10^{-4}$) with dimensionless coherence length of $l_B=25.69$;
(b) $\frac{\C}{\omega_B\omega_C}=102.9936$,
at the border of weak and strong correlations of the BEC,
i.e., when the BEC coherence length is $l_B=1$ (the depletion is $\rho_B\approx 0.1715$);
and (c) $\frac{\C}{\omega_B\omega_C}=810.0779$,
when the BEC in $50\%$ fragmented (the coherence length is $l_B=\frac{1}{2\sqrt{2}}\approx 0.3535$).
See Fig.~\ref{F1} for the widths of the respective densities and the text for further discussion.
The quantities shown are dimensionless.}
\label{F3}
\end{figure}

In Figs.~\ref{F1}, \ref{F2}, and \ref{F3},
we follow properties of the BEC and the cavity as a function of the coupling parameter $\C$.
We remind that all properties are analytically expressed in closed form as a function of the parameters in the Hamiltonian. 
We begin our analysis with the simplest quantity,
the density.
To recall,
for the bosons the density per particle is $\frac{\rho_B(x)}{N}=
\left(\frac{\alpha_B+C_{1,0}}{\pi}\right)^{\frac{1}{2}}
e^{-\left(\alpha_B+C_{1,0}\right) x^2}$
and for the cavity the density is
$\rho_C(q) = \left(\frac{\alpha_C+C'_{0,1}}{\pi}\right)^{\frac{1}{2}}$.
Fig.~\ref{F1} depicts the widths of the densities
$\frac{1}{\sqrt{\alpha_B+C_{1,0}}}$ and $\frac{1}{\sqrt{\alpha_C+C'_{0,1}}}$, respectively,
as a function of the coupling parameter.

Increasing the coupling,
the BEC barely changes its density in real space 
over five decades of the coupling parameter $\frac{\C}{\omega_B\omega_C}$.
The cavity wavepacket, on the other hand, exhibits different behavior.
Up to about $\frac{\C}{\omega_B\omega_C} \approx 1$ the width is almost
unchanged and past this value the cavity wavepacket is seen to expand in Fock space.
Clearly, something is happening for the cavity at about  $\frac{\C}{\omega_B\omega_C} \approx 1$.
We will see, in the appendix, that this value marks the breakdown for the cavity mode 
of the mean-field theory for the BEC-cavity system.
Interestingly,
if one were to neglect the dipole self-interaction term in the Hamiltonian (\ref{HAM_BC},\ref{HAM_BC_JAC1}), 
the value $\frac{\C}{\omega_B\omega_C}=1$
would have been the upper allowed bound for the coupling parameter.

Analysis of the respective widths for strong coupling
provides asymptotic expressions.
The dependence of the BEC width, to leading order in $\C$, 
in the large coupling-parameter limit $\frac{\C}{\omega_B\omega_C} \gg 1$ is
$\frac{1}{\sqrt{\alpha_B+C_{1,0}}} \to 
\sqrt{\frac{1+\frac{1}{N}\left[\left(1+\frac{\omega_C}{\omega_B}\right)
\frac{\omega_B\omega_C}{\C}-1\right]}{m\omega_B}}$.
Hence,
when the BEC is coupled to the cavity it becomes only slightly narrower,
down to the factor of $\sqrt{1-\frac{1}{N}}$ than in the absence of coupling.
For $N=100$ bosons the difference is already very small
and amounts to half a percent.
Such a minor effect on the density of the BEC
can be explained by
recalling that only a single out of $N$ degrees-of-freedom, the bosons center-of-mass,
is coupled to the cavity mode.
Similarly,
the dependence of the cavity width, to leading order in $\C$, 
in the large coupling-parameter limit $\frac{\C}{\omega_B\omega_C} \gg 1$ is
$\frac{1}{\sqrt{\alpha_C+C'_{0,1}}} \to \sqrt{\frac{1}{m\omega_C}\frac{G}{\omega_B\omega_C}}$,
and seen to grow with the coupling parameter.
Both asymptotic functions are plotted in Fig.~\ref{F1} for completeness to guide the eye.
What other implications
accompany the saturation of the BEC density in real space and
the broadening of the cavity density in Fock space?

Fig.~\ref{F2} follows
the depletion $\rho_B$ of the BEC
and the analogously-computed $\rho_C$ for the cavity mode
as a function of the coupling parameter $\C$.
The cavity depletion is also 
the Schmidt-decomposition parameter quantifying the von Neumann 
entanglement entropy of the BEC and the cavity,
as it satisfies $\rho_{SD}^2=\rho_C$.
As noted above,
the von Neumann 
entanglement entropy of the coupled BEC and cavity, Eq.~(\ref{SD_ENT_ENT}),
follows $\rho_{SD}^2$ monotonously,
making the depletion
of the cavity a one-to-one representation of the entanglement entropy.
Another interesting property
is that the depletion of the cavity $\rho_C$ does not depend on the number of bosons $N$
in the BEC.

One sees from Fig.~\ref{F2}a that the depletion of
the BEC grows with the coupling parameter $\C$,
turning the BEC more and more correlated.
Let us pin point a few values of interest.
For $\frac{\C}{\omega_B\omega_C}=1$ the depletion is quite low,
$\rho_B=3.784 \cdot 10^{-4}$.
As mentioned above,
this is the value of the coupling parameter for which the cavity description
at the mean-field level breaks down.
The above-discussed border between weak and strong correlations of the bosons
is reached for $\frac{\C}{\omega_B\omega_C}=102.9936$. 
To remind, here the dimensionless coherence length is $\l_B=1$
and the corresponding depletion $\rho_B =\frac{\sqrt{2}-1}{\sqrt{2}+1}\approx 0.1715$.
Finally, for $\frac{\C}{\omega_B\omega_C}=810.0779$
the BEC becomes 50\% fragmented,
i.e., the depletion is $\rho_B=0.5$ and
the natural occupation numbers per particle become $0.5, 0.25, 0.125, 0.0625, \ldots$.
As can be inferred from Fig.~\ref{F2},
increasing the coupling parameter the degree of fragmentation further increases.

Interestingly,
for non-interacting bosons
the depletion $\rho_B$ of the BEC
depends on the mass $m$ of a boson only through the functional dependence of the 
coupling parameter $\C=g \sqrt{\frac{2\omega_CN}{m}}$.
This implies that
BECs with different boson masses
possess the same degree of fragmentation for a fixed coupling parameter $\C$.
Of course, the larger the mass the narrower is the fragmented BEC.
All in all,
we find for the BEC as a result of its coupling to the cavity,
that the cavity induces effective long-range interaction,
which in turn governs rich correlations and fragmentation, 
between the otherwise non-interacting bosons.

Turning to the cavity, its depletion $\rho_C$ also grows with $\C$.
But, unlike the depletion of the BEC,
$\rho_C$ eventually saturates, see Fig.~\ref{F3}b.
Accordingly,
the von Neumann 
entanglement entropy (\ref{SD_ENT_ENT}) saturates as well. 
The asymptotic value
of the depletion for large coupling parameter $\frac{\C}{\omega_B\omega_C} \gg 1$ 
can be readily be evaluated and is given by
$\rho_C \to \frac{\sqrt{1+\frac{\omega_C}{\omega_B}}-1}{\sqrt{1+\frac{\omega_C}{\omega_B}}+1}$.
Using (\ref{VAC_LC}),
this implies a simple and appealing result for the saturation value of the cavity dimensionless coherence length,
$l_C \to \sqrt{\frac{\omega_B}{\omega_C}}$.
Thus, the asymptotic value of depletion is governed
by the ratio of the cavity and trap frequencies.
For $\omega_C=0.1$ one gets a low value for the asymptotic depletion $\rho_C \to 0.02382$,
or, the cavity dimensionless coherence length goes to $l_C \to \sqrt{10}\approx 3.162$,
i.e., the cavity is weakly correlated.
Hence,
to get a larger degree of depletion a larger cavity frequency is required.
For instance,
for $\omega_C=10.0$ one obtains $\rho_C \to 0.5366$, see Fig.~\ref{F3}b.
The corresponding coherence length becomes $l_C\to \sqrt{\frac{1}{10}}\approx 0.3162$,
meaning that the cavity is now strongly
correlated as a result of its coupling to the bosons.

The properties of the bosons and the cavity found above are complementary.
The density of the bosons is practically unchanged and saturates,
whereas its depletion increasingly grow with the coupling parameter.
On the other hand,
the density of the cavity increasingly broadens,
whereas its depletion saturates.
What are the resulting implications on the correlation functions and
coherence lengths
of the bosons and the cavity?

For the cavity, the saturation of the depletion $\rho_C$ means, as we have seen above,
saturation of the correlation length $l_C = \sqrt{\frac{\alpha_C+C'_{0,1}}{-C'_{0,1}}}$ [Eq.~(\ref{VAC_LC})].
But, the growth of the width of the density, $\frac{1}{\sqrt{\alpha_C+C'_{0,1}}}$, see Fig.~\ref{F1},
implies that the cavity correlation function 
$g_C(q,q') = e^{-\frac{-C'_{0,1}}{4}\left(q-q'\right)^2}$
must decay slower to precisely generate the saturation of the coherence length.
For the bosons the situation and arguments are reversed.
Recall that the width of the density, $\frac{1}{\sqrt{\alpha_B+C_{1,0}}}$, 
saturates to roughly the value of the uncoupled BEC,
see Fig.~\ref{F1},
while the depletion $\rho_B$ keeps growing, see Fig.~\ref{F2}a.
This implies that the bosons correlation function
$g_B(x,x') = e^{-\frac{-C_{1,0}}{4}\left(x-x'\right)^2}$ [Eq.~(\ref{g1_BEC})]
must decay faster to generate the decrease of the coherence length 
$l_B = \sqrt{\frac{\alpha_B+C_{1,0}}{-C_{1,0}}}$ [Eq.~(\ref{BEC_LB})]
when the BEC becomes more fragmented.

Fig.~\ref{F3} shows  
the first-order spatial correlation function of the BEC, $g_B(x,x')$,
for the above-discussed three coupling parameters.
Fig.~\ref{F3}a is
when the mean-field theory breaks down for the cavity ($\frac{\C}{\omega_B\omega_C}=1$), also see the appendix,
and the bosons are weakly depleted.
Fig.~\ref{F3}b is
at the border of weak and strong correlations of the bosons ($\frac{\C}{\omega_B\omega_C}=102.9936$).
Finally, Fig.~\ref{F3}c is
when the BEC becomes 50\% fragmented ($\frac{\C}{\omega_B\omega_C}=810.0779$).
The decay of $g_B(x,x')$
and the decrease of $l_B$
for the essentially-constant BEC density
are evident.
The respective values of the dimensionless coherence lengths are
$l_B=25.69$, $l_B=1$, and $l_B=\frac{1}{2\sqrt{2}}\approx 0.3535$.

Before concluding,
we expand the analysis of the BEC-cavity system 
and touch upon the so-called infinite-particle-number limit.
At the infinite-particle-number limit of trapped bosonic systems,
the interaction parameters are held fixed while the number of bosons is increased to infinity.
One inquires on the relation between the many-body and mean-field theories for the ground state of interacting bosons in this limit \cite{INF1,INF2,INF3,INF4}.
Whereas the exactness of the mean-field theory
for the energy per particle, density per particle,
and 100\% condensate fraction 
has been proved \cite{INF1,INF2},
other properties,
such as variances of observables per particle,
require a many-body theory
even at the infinite-particle-number limit
when the bosons are 100\% condensed \cite{INF3}.

What are the relevant interaction parameters,
to be held fixed when performing the infinite-particle-number limit,
in the BEC-cavity system?
These parameters are drawn from the corresponding mean-field theory
which is given in the appendix in our case.
For the bosons,
it would be the standard interaction parameter $\lambda_B(N-1)$ in which the boson-boson interaction strength
is inversely proportional to the number of bosons.
For non-interacting bosons,
the interaction parameter is obviously zero for any number of bosons.
For the cavity, there are two options for such a parameter,
when reviewing the mean-field equations (\ref{MF_EQs}) in the appendix.
The first would be the analogous parameter $gN$, i.e.,
when the BEC-cavity coupling strength is inversely proportional to the number of bosons. 
The second option would be $g\sqrt{N-1}$,
that is,
when $g$ is inversely proportional to the square root of the number of bosons.
Clearly, in the second case,
the coupling constant $\C=g\sqrt{\frac{2\omega_CN}{m}}$ itself
is held fixed at the infinite-particle-number limit.

It is instructive to analyze what is happening at this limit to the fragmentation of the bosons,
correlations of the cavity,
and entanglement between the bosons and cavity mode.
The next data point in this limit, after the lead example with $N=100$ bosons,
would be $N=1000$ bosons, see Fig.~\ref{F2}b.
The degree of depletion and with it the level of fragmentation as one would anticipate decrease.
Performing the infinite-particle-number limit 
to the reduced one-particle density matrix of the BEC (\ref{1RDM_BEC}),
we find that the bosons become 100\% condensed,
as is expected for single-species and mixtures of bosons \cite{BB_HIM,INF2,INF5}.
On the other hand,
the correlation properties of the cavity 
and the BEC-cavity entanglement, see (\ref{1RDM_CAV}) and (\ref{BC_WF_ENT_SD}),
depend on whether one performs the infinite-particle-number limit
with $gN$ held fixed or with $g\sqrt{N-1}$ fixed.
In the former case,
the cavity becomes uncorrelated and the entanglement vanishes.
Conversely,
in the latter case the cavity mode remains correlated and
the entanglement between the bosons and the cavity mode persists
at the infinite-particle-number limit and does not vanish.
This is a good place to conclude the present investigation.

\section{Summary and outlook}\label{SEC7}

The present work devises and solves analytically a many-body model
for a Bose-Einstein condensate in a single-mode cavity,
within the dipole approximation.
The analyses and outcomes of the theory allows us to investigate
the impact of the cavity on the
bosons and vice versa on an equal footing.
The bosons and the cavity mode are treated on an equal footing.
After obtaining the energy and wavefunction of the 
coupled BEC-cavity system,
properties are computed explicitly.
The questions of how the cavity impacts the BEC,
how the BEC influences the cavity,
and how the two are entangled
are asked and answered within the model.
To this end, the reduced one-particle density matrices
are diagonalized,
and their eigenvalues associated with the depletion
or correlations.
It is useful to define a dimensionless
coherence length
and use it to assess how weak or strong correlations are.
For the BEC, the coherence length
compares the spatial decay of
the first-order correlation function and size of the density.
For the cavity,
it is an analogously-defined measure for the strength of correlations
generated by the coupling to the BEC.
Finally,
the access to all natural occupation numbers of the cavity reduced one-particle density
matrix allows one to obtain a closed-form expression for the von Neumann
entanglement entropy of the BEC-cavity system.
The latter has also been evaluated directly by Schmidt-decomposing
the BEC-cavity wavefunction.
In this capacity,
the BEC and the cavity can be a highly-entangled many-body system,
see below.

Application of the theory includes a specific case with fundamental interest of its own,
harmonically-trapped non-interacting bosons in a cavity.
We have evaluated all the above quantities in closed form and analyzed them,
finding that the bosons and the cavity
develop correlations in a complementary manner
when increasing the coupling between them. 
Whereas the density of the BEC in real space saturates with $\frac{\C}{\omega_B\omega_C}$,
the analogously-defined cavity density broadens in Fock space.
Side by side,
the cavity depletion saturates,
and hence the BEC-cavity entanglement entropy does.
The saturation value of the depletion and entanglement is controlled by and increases with 
the ratio of the cavity and trap frequencies. 
With increasing the coupling parameter,
the BEC becomes strongly correlated and eventually increasingly more fragmented.
The phenomenon constitutes fragmentation in a single trap \cite{Single_trap1,Single_trap2} of otherwise ideal bosons,
where the cavity induces long-range interaction between them.
Similarly to ordinary fragmentation of interacting single-species bosons in a trap potential,
the fragmentation of non-interacting bosons in a cavity at constant coupling $\frac{\C}{\omega_B\omega_C}$
decreases with the number of bosons. 

Finally, to shed further light on the role of entanglement between the bosons and the cavity
and as a complementary investigation,
we solved analytically also the mean-field equations for the description of the cavity and the BEC. 
The analysis is collected in the appendix.
The mean-field theory completely neglects the entanglement
between the bosons and the cavity, as well as correlations between the bosons themselves.
Using the two analytical solutions, the many-body and the mean-field,
we identified for the above case of non-interacting bosons in a cavity
when the mean-field description breaks down as a function of the coupling parameter
for the simplest quantity, the density.
For the cavity the breakdown occurs at about 
$\frac{\C}{\omega_B\omega_C} \approx 1$
and for the BEC at roughly
$\frac{\C}{\omega_B\omega_C} \approx \sqrt{N}$,
which matches the numbers of degrees-of-freedom
in the cavity and the BEC.

The present model of bosons interacting by harmonic forces in a cavity
is not a replacement of more realistic interactions,
whether long ranged or short ranged.
Rather, its usage would be to develop further intuition
for more realistic systems.
It would be instructive to extend the model itself,
for instance by including dissipation
and driving to the cavity mode or,
independently,
accounting for internal structure of the bosons.
One could envision additional applications and extensions.
For instance,
drawing from \cite{Cavity_ROT,Floquet_HIM},
investigating the exchange of angular momentum
between a BEC and a cavity would be interesting.
Another extension,
in the spirit of \cite{Braun02},
would be for the correlations induced 
in two cavity modes when coupled to a BEC,
or, the entanglement generated between two BECs when coupled to a single cavity mode,
and so on.
We believe there is a vast ground
for further developments
based on the present exactly-solvable
model of a BEC in a cavity.

\section*{Acknowledgements}

This research was supported by the Israel Science Foundation
(Grant No. 1516/19).

\appendix

\section{On the mean-field solution of a Bose-Einstein condensate in a cavity}\label{SEC8}

The mean-field theory of Bose-Einstein condensates, Gross-Pitaevskii equation, has been highly popular
due to its simplicity and transparency \cite{rev1,rev2}.
The exactness of mean-field theory at the limit of an infinite number of bosons
and fixed interaction parameter has been mathematically rigorously proved
for the energy per particle, density per particle,
and 100\% condensate fraction in the ground state \cite{INF1,INF2}.
We note that for other properties,
such as variances of observables per particle,
a many-body theory is required \cite{INF3}
even in the above limit of an infinite number of particles
when the bosons are 100\% condensed.

In this appendix, we derive the mean-field equations
for the BEC-cavity system,
discuss some of their properties,
and solve them 
for the specific case of non-interacting bosons in a cavity
investigated in the main text at the many-body level of theory.
It is well known that the mean-field approximation, being a product state,
does not describe correlations in BECs
and cannot account for entanglement in mixtures of BECs.
We do not expect it to do so here,
in the BEC-cavity system.
Below,
comparing the analytical solutions at the mean-field and many-body
levels of theory allows us to identify and quantify
the consequences of using mean-field theory
for the BEC-cavity system.

The Hamiltonian (\ref{HAM_BC},\ref{12Q}) reads in the laboratory frame
\beqn\label{HAM_CARz}
& &
H = \sum_{j=1}^N \left[ -\frac{1}{2m} \frac{\partial^2}{\partial x_j^2} + 
V(x_j) + \frac{g^2}{\omega_C} x_j^2\right] + \sum_{1 \le j < k}^{N} \left[W(x_j-x_k) + \frac{2g^2}{\omega_C} x_jx_k\right] + \nonumber \\
& & 
+ \left(-\frac{1}{2m} \frac{\partial^2}{\partial q^2} + \frac{1}{2} m {\omega_C}^2 q^2\right) +
g \sqrt{2m\omega_C} \sum_{j=1}^{N} x_jq. \
\eeqn
There are three contributions to the many-body Hamiltonian (\ref{HAM_CARz}) resulting from the coupling of the BEC and the cavity.
The first is a one-body harmonic potential,
the second is a long-range interaction-like two-body term,
and the third is the two-body term coupling the bosons and cavity coordinates.

Take the mean-field ansatz
\beqn\label{MF_ANZ}
\Phi^{MF}(x_1,\ldots,x_N,q) = \phi(x_1) \cdots \phi(x_N) \varphi(q) \
\eeqn
and sandwich it with the Hamiltonian (\ref{HAM_CARz}).
The mean-field energy functional
is then obtained and reads
\beqn\label{MF_EF}
& &
E^{MF} = N \int dx \phi^\ast(x) \left[ -\frac{1}{2m} \frac{\partial^2}{\partial x^2} + V(x) + \frac{g^2}{\omega_C} x^2\right] \phi(x) + \nonumber \\
& &
+ \frac{N\left(N-1\right)}{2} \int dx dx' |\phi(x)|^2 |\phi(x')|^2 \left[W(x-x') + \frac{2g^2}{\omega_C} xx'\right] + \nonumber \\
& &
+ \int dq \varphi^\ast(q) \left(-\frac{1}{2m} \frac{\partial^2}{\partial q^2} + \frac{1}{2} m {\omega_C}^2 q^2\right)
\varphi(q) + \nonumber \\
& &
+ N g \sqrt{2m\omega_C} \int dx dq |\phi(x)|^2 |\varphi(q)|^2 x q. \
\eeqn
Equating the variations of (\ref{MF_EF}) with respect to $\phi^\ast(x)$ and $\varphi^\ast(q)$ to zero one finds
\beqn\label{MF_EQs}
& &
\left\{-\frac{1}{2m} \frac{\partial^2}{\partial x^2} + V(x) + \frac{g^2}{\omega_C} x^2 +
\left(N-1\right) \left[\int dx' |\phi(x')|^2 W(x-x') + \frac{2g^2}{\omega_C} x \int dx' |\phi(x')|^2x'\right] + \right. \nonumber \\
& & 
\qquad \left. + g \sqrt{2m\omega_C} x \int dq |\varphi(q)|^2 q\right\} \phi(x) = \mu_B \phi(x),
\nonumber \\
& &
\left\{-\frac{1}{2m} \frac{\partial^2}{\partial q^2} + \frac{1}{2} m {\omega_C}^2 q^2 +
N g \sqrt{2m\omega_C} q \int dx |\phi(x)|^2 x \right\}\varphi(q)=\mu_C \varphi(q), \
\eeqn
where $\mu_B$ and $\mu_C$ are the corresponding chemical potentials
(Lagrange multipliers ensuring normalization of the orbitals) for the BEC and the cavity mode.
These are the coupled mean-field equations for the BEC-cavity system.

Let us analyze the mean-field equations (\ref{MF_EQs}) and their prospected solutions more closely.
We focus first on the non-linear coupling terms where the cavity
density $|\varphi(q)|^2$ appears in the equation for the bosons,
and the bosons density $N|\phi(x)|^2$ appears in the equation for the cavity.
Generally, the mean-field solution is expected to be symmetric
when the many-body solution is symmetric
(a remark on the possibility of symmetry-broken mean-field solutions appears below).
The one-body potential of the cavity mode is harmonic and obviously a symmetric function of the coordinate $q$.
Thus, when the trap of the bosons is spatially reflection symmetric, $V(x)=V(-x)$,
the solutions $\phi(x)$ and $\varphi(q)$ are even functions of the $x$ and $q$ coordinates, respectively.
Consequently, the overlap integrals of the densities 
$|\varphi(q)|^2$ and $N|\phi(x)|^2$ vanish
and (\ref{MF_EQs}) simplifies considerably,
\beqn\label{MF_SYM_EQs}
& &
\left\{-\frac{1}{2m} \frac{\partial^2}{\partial x^2} + V(x) + \frac{g^2}{\omega_C} x^2 +
\left(N-1\right)\int dx' |\phi(x')|^2 W(x-x')\right\} \phi(x) = \mu_B \phi(x),
\nonumber \\
& &
\left\{-\frac{1}{2m} \frac{\partial^2}{\partial q^2} +
\frac{1}{2} m {\omega_C}^2 q^2\right\}\varphi(q)=\mu_C \varphi(q). \
\eeqn
The BEC-cavity two mean-field equations (\ref{MF_SYM_EQs}) are found to decouple from each other,
i.e.,
the BEC solution $\phi(x)$ and the cavity solution $\varphi(q)$
are determined independently.
On top of that,
the only dependence on the coupling 
appearing at the many-body level (\ref{HAM_CARz}) 
and left at the mean-field level of theory
is the one-body 
harmonic-potential term in the BEC equation (\ref{MF_SYM_EQs}).
The solution $\varphi(q)$ for the cavity is simply the bare harmonic-oscillator eigenfunction.

\begin{figure}[!]
\begin{center}
\hglue -1.6 truecm
\includegraphics[width=0.43\columnwidth,angle=-90]{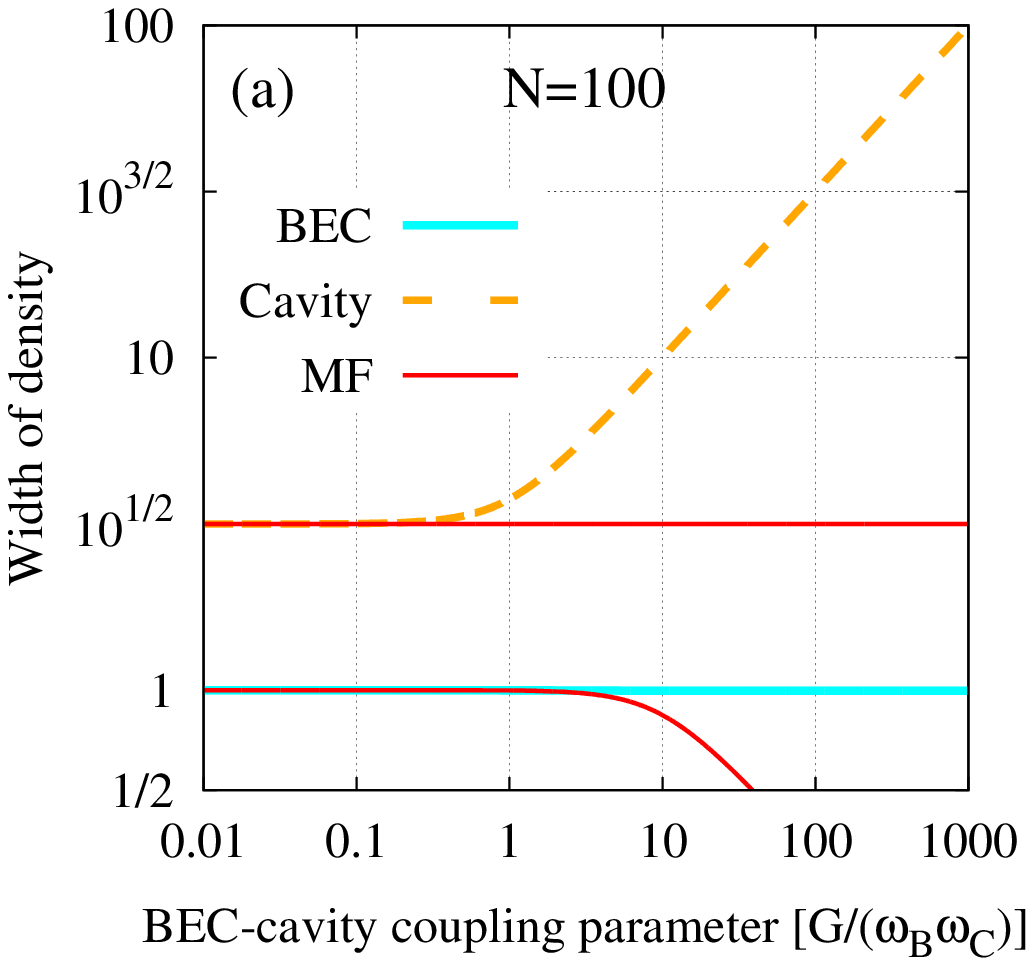}
\hglue -2.8 truecm
\includegraphics[width=0.43\columnwidth,angle=-90]{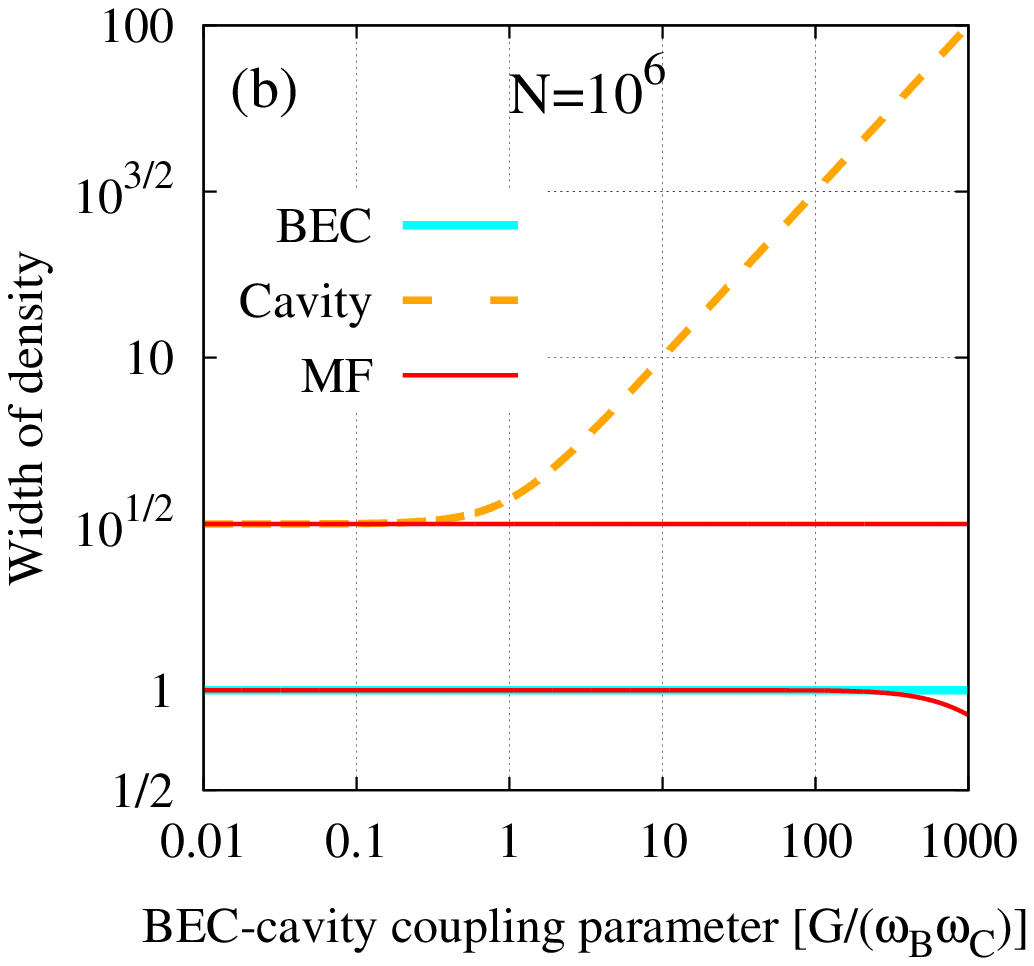}
\end{center}
\vglue 0.15 truecm
\caption{Mean-field widths
of the bosons density per particle and the analogously-defined
cavity density.
The many-body results are plotted for comparison.
(a) $N=100$ and (b) $N=10^6$ 
non-interacting bosons of mass $m=1.0$ in a trap ($\omega_B=1.0$)
are coupled to a single-mode quantum light in a cavity ($\omega_C=0.1$).
Qualitative differences appear for the bosons
for coupling parameters larger than $\frac{\C}{\omega_B\omega_C} \approx \sqrt{N}$,
and for the cavity for coupling
parameters greater than $\frac{\C}{\omega_B\omega_C} \approx 1$.
See the text for further analysis and discussion.
The quantities shown are dimensionless.}
\label{F4}
\end{figure}

Let us consider an application,
non-interacting bosons held in harmonic potential and coupled to a cavity.
The first system is taken to consist $N=100$ bosons,
like in the main text,
and the second is assumed to contain a much larger number of bosons, $N=10^{6}$.
All other parameters are same as in Fig.~\ref{F1},
namely the trapping frequency is $\omega_B=1.0$, the cavity frequency is $\omega_C=0.1$,
and the mass of a bosons is $m=1.0$.
Then, Eq.~(\ref{MF_SYM_EQs}) further simplifies
since also the BEC equation becomes linear
and the combined one-body potential turns harmonic,
$\frac{1}{2}m\left(\omega^2_B+\frac{2g^2}{m\omega_C} \right)x^2 =
\frac{1}{2}m\omega^2_B\left[1+\frac{1}{N}\left(\frac{G}{\omega_B\omega_C}\right)^2\right]x^2$.
Fig.~\ref{F4} depicts the width of the density per
particle of the bosons, $\frac{1}{\sqrt{m\omega_B\sqrt{1+\frac{1}{N}\left(\frac{G}{\omega_B\omega_C}\right)^2}}}$,
and the width of the analogous cavity density, $\frac{1}{\sqrt{m\omega_C}}$,
at the mean-field level of theory.
The widths of the respective densities
at the many-body level of theory are plotted too for comparison.

It is found and shown that
the mean-field theory does not describe the growth of the cavity density
from about $\frac{\C}{\omega_B\omega_C} \approx 1$,
irrespective to the number of bosons.
The mean-field width is constant
and the many-body width behaves for $\frac{\C}{\omega_B\omega_C} \gg 1$
as $\frac{1}{\sqrt{\alpha_C+C'_{0,1}}} \to \sqrt{\frac{1}{m\omega_C}\frac{G}{\omega_B\omega_C}}$.
We remind that the many-body solution for the cavity also does not explicitly depend on $N$,
but only implicitly via $G$.
Hence, there is no way to reconcile or mitigate the differences between the
mean-field and many-body treatments of the cavity,
except for attributing the matter to the absence of entanglement
between the BEC and the cavity at the mean-field level of theory.
Side by side,
the mean-field theory predicts a shrinking of the bosons density
from about $\frac{\C}{\omega_B\omega_C} \approx \sqrt{N}$,
i.e., the larger is the BEC the stronger is the coupling needed to get
deviation between the mean-field and many-body theories.
We recall that the many-body solution only weakly depends on $N$,
with width
$\frac{1}{\sqrt{\alpha_B+C_{1,0}}} \to 
\sqrt{\frac{1+\frac{1}{N}\left[\left(1+\frac{\omega_C}{\omega_B}\right)
\frac{\omega_B\omega_C}{\C}-1\right]}{m\omega_B}}$,
and more so it saturates for $\frac{\C}{\omega_B\omega_C} \gg 1$ at nearly the uncoupled value.
All in all,
the mean-field solution to non-interacting bosons in a cavity
cannot describe even qualitatively
the width of the density
of the BEC, for coupling
parameters larger than $\frac{\C}{\omega_B\omega_C} \approx \sqrt{N}$,
and even more severely for the cavity,
for coupling
parameters greater than 
$\frac{\C}{\omega_B\omega_C} \approx 1$.

To conclude, three complementary remarks are in place:
1) The same symmetry argument leading to (\ref{MF_SYM_EQs}) would hold true for the more general yet separable ansatz
$\Phi(x_1,\ldots,x_N,q) = \Theta(x_1,\ldots,x_N) \varphi(q)$,
leading to two decoupled equations, one for the bosons and one for the cavity.
The solution $\varphi(q)$ for the cavity is again simply the bare harmonic-oscillator eigenfunction.
2) The vanishing of coupling at the mean-field level of theory for the
symmetry-preserving solutions $\phi(x)$ and $\varphi(q)$
makes the mean-field equations (\ref{MF_EQs}) intriguing,
in the manner whether symmetry-broken mean-field solutions for the BEC-cavity system exist.
Indeed, even though the boson-boson interaction $W(x-x')$
is spatially reflection symmetric,
the corresponding non-linear term in the mean-field equation for the bosons can lead
to symmetry-broken solutions for sufficiently strong non-linear interactions.
When this is the case,
coupling between the solutions of the BEC and the cavity
at the mean-field level of theory is restored.
3) Fig.~\ref{F4} presents the mean-field solutions for non-interacting bosons in a cavity.
We note for completeness that the analytical solution of
the mean-field equations (\ref{MF_EQs},\ref{MF_SYM_EQs})
can be obtained also for interacting bosons in the cavity (\ref{HAM_BC_JAC1}),
similarly to \cite{HIM_COHEN,BB_HIM}.

\end{document}